\documentclass[12pt]{iopart}
\usepackage{graphicx}
\usepackage{color}
\usepackage{iopams}

\newcommand{\Otwo}{{O$_{2}$ }}
\newcommand{\Ntwo}{{N$_{2}$ }}

\begin{document}

\title[Oxygen super rotors]{Coherent spin-rotational dynamics of oxygen super rotors}

\author{Alexander A. Milner, Aleksey Korobenko and Valery Milner}

\address{Department of Physics \& Astronomy, University of British Columbia, 2036 Main Mall, Vancouver, BC, Canada V6T 1Z1.}

\ead{vmilner@phas.ubc.ca}

\begin{abstract}
We use state- and time-resolved coherent Raman spectroscopy to study the rotational dynamics of oxygen molecules in ultra-high rotational states. While it is possible to reach rotational quantum numbers up to $N \approx 50$ by increasing the gas temperature to 1500~K, low population levels and gas densities result in correspondingly weak optical response. By spinning \Otwo molecules with an optical centrifuge, we efficiently excite extreme rotational states  with $N\leqslant 109$ in high-density room temperature ensembles. Fast molecular rotation results in the enhanced robustness of the created rotational wave packets against collisions, enabling us to observe the effects of weak spin-rotation coupling in the coherent rotational dynamics of oxygen. The decay rate of spin-rotation coherence due to collisions is measured as a function of the molecular angular momentum and explained in terms of the general scaling law. We find that at high values of $N$, the rotational decoherence of oxygen is much faster than that of the previously studied non-magnetic nitrogen molecules. This may suggest a different mechanism of rotational relaxation in paramagnetic gases.
\end{abstract}

\maketitle

Rotational spectroscopy of molecular oxygen, one of the most abundant molecules in the earth's atmosphere, is key to many studies in physics and chemistry, from atmospheric science and astronomy\cite{Slanger97} to thermochemistry and combustion research\cite{Martinsson96, Thumann97, Seeger09, Miller11a}. Among simple diatomic molecules, \Otwo stands out because of its nonzero electron spin ($S=1$) in the ground electronic state, $X^{3}\Sigma ^{-}_{g}$. The interaction between the spins of the two unpaired electrons and the magnetic field of the rotating nuclei results in the spin-rotation (SR) coupling on the order of a few wave numbers, which grows with increasing nuclear rotation quantum number $N$\cite{HerzbergBook}. This coupling of the electron magnetism with molecular rotation, readily controllable with laser light, offers new opportunities for controlling molecular dynamics in external magnetic fields\cite{Gershnabel11}.

\begin{figure}[b]
\centering
\includegraphics[width=.9\columnwidth]{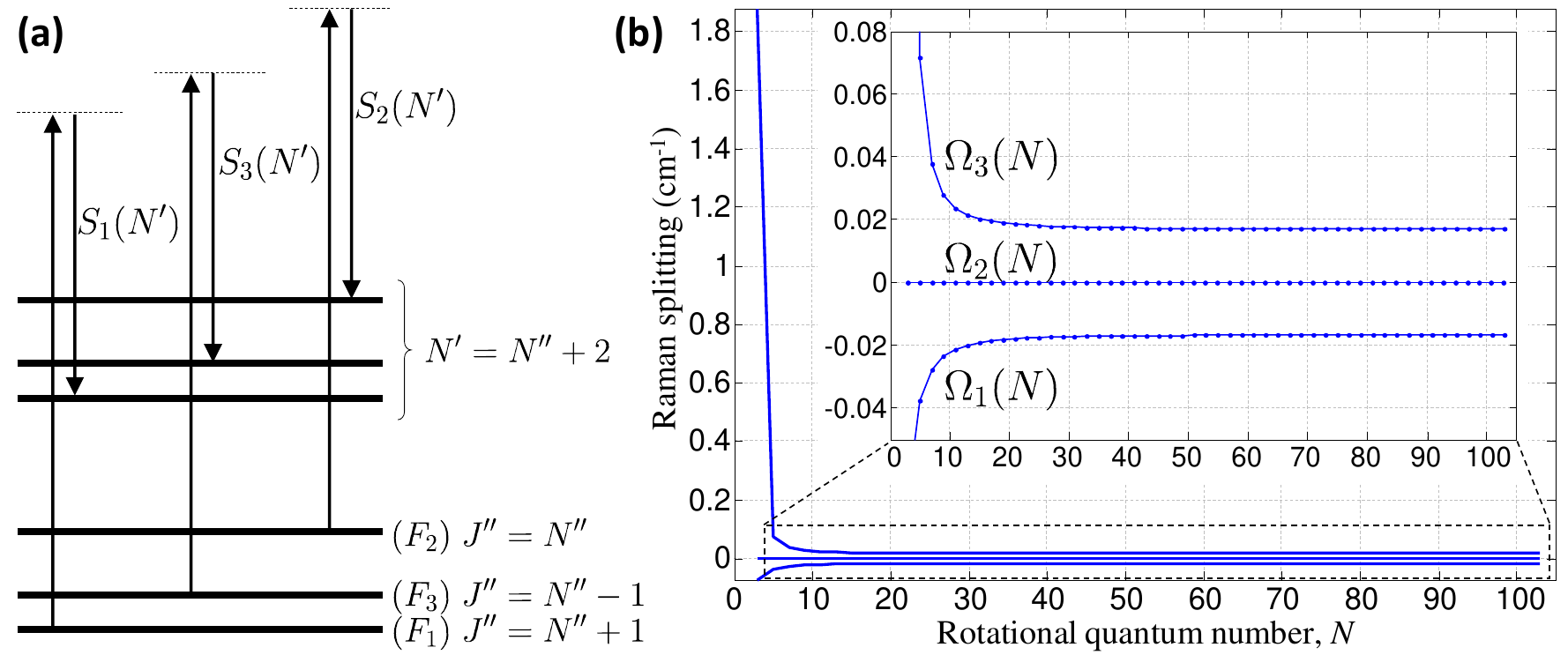}
\caption{(\textbf{a}): Spin-rotational splitting of two rotational levels of oxygen, $N''$ and $N'=N''+2$. Each level is split into three sub-levels with energies $F _{k}, k=1,2,3$ for the total angular momentum $J=N+1,N,N-1$, respectively. Three strongest Raman transitions (out of the total six allowed by the selection rules) corresponding to the $S(N')$ branch are shown and labeled according to the participating $J$-states. (\textbf{b}): Dependence of the three Raman frequencies ($\Omega_{k}$ for $S_{k}$ line) on the rotational quantum number.}
\label{Fig-Splitting}
\end{figure}
SR coupling splits each rotational level in three (Fig.\ref{Fig-Splitting}(\textbf{a})), with the total angular momentum $J=N, N\pm1$ \cite{HerzbergBook}. The energy splitting, originally observed by Dieke and Babcock in 1927\cite{Dieke27} and later calculated by Kramers\cite{Kramers29} and Schlapp\cite{Schlapp37}, is currently known with a very high degree of accuracy\cite{Yu12}. Though routinely observed in the frequency domain with the methods of microwave spectroscopy\cite{Beringer51, Miller53}, to the best of our knowledge, the spin-rotation splitting and the associated with it SR dynamics have not been studied in the time domain.

Time-resolved coherent Raman scattering is a common tool of choice for analyzing the dynamics of molecular rotation. It has been successfully used for the precision thermometry of flames\cite{Seeger09}, the studies of collisional decoherence in dense gas media\cite{Miller11, Kearney13}, and recently by our own group for the detection and study of molecular super rotors\cite{Korobenko14a, Milner14a}. Because of the spin-rotation coupling, any $N \rightarrow (N+2)$ Raman transition in oxygen consists of six separate lines belonging to one $Q$, two $R$ and three $S$ branches with $\Delta J=0,1$ and 2, respectively. The strength of both $Q$ and $R$ branches drops quickly with increasing $N$ and becomes negligibly small at $N>5$ \cite{Berard83}. The frequency difference between the lowest $R$ lines, $R(1)$ and $R(3)$, is about 2 cm$^{-1}$. Their interference in the time domain results in the oscillations with a period of $\approx17$ ps, which has been recently observed experimentally\cite{Miller11}. On the other hand, the three stronger $S$ branches shown in Fig.\ref{Fig-Splitting}(\textbf{a}) are split by less than 0.05 cm$^{-1}$ (Fig.\ref{Fig-Splitting}(\textbf{b})), which corresponds to the oscillation period of about 600 ps. This time scale is much longer than the collisional decoherence time of the thermally populated rotational levels at ambient pressure\cite{Berard83, Millot92}, explaining why no spin-rotational dynamics has been seen for $N>3$ in the time-resolved experiments\cite{Miller11}.

In this work we employ the technique of an optical centrifuge\cite{Karczmarek99, Villeneuve00} to excite oxygen molecules to ultra-high angular momentum states, reaching rotational quantum numbers as high as $N=109$. Similarly to our previous studies of nitrogen super rotors\cite{Milner14a}, we observe that the life time of rotational coherence in oxygen becomes substantially longer at high $N$'s, making the detection of spin-rotational oscillations possible even at the pressure of 1 atmosphere. By lowering the pressure, we observe SR dynamics in the broad range of angular momentum, $3\leqslant N \leqslant 109$. We measure the decay rate of the spin-rotation oscillations due to O$_{2} - $O$_{2}$ collisions and analyze its dependence on $N$ using the energy corrected sudden (ECS) model of rotational energy transfer. In contrast to other experimental methods, our ability to vary the speed of molecular rotation without changing the temperature of the gas allows us to reach the adiabatic limit of rotational relaxation, when the period of molecular rotation becomes shorter than the collision time\cite{Milner14a}.
\begin{figure}[t]
\centering
\includegraphics[width=.9\columnwidth]{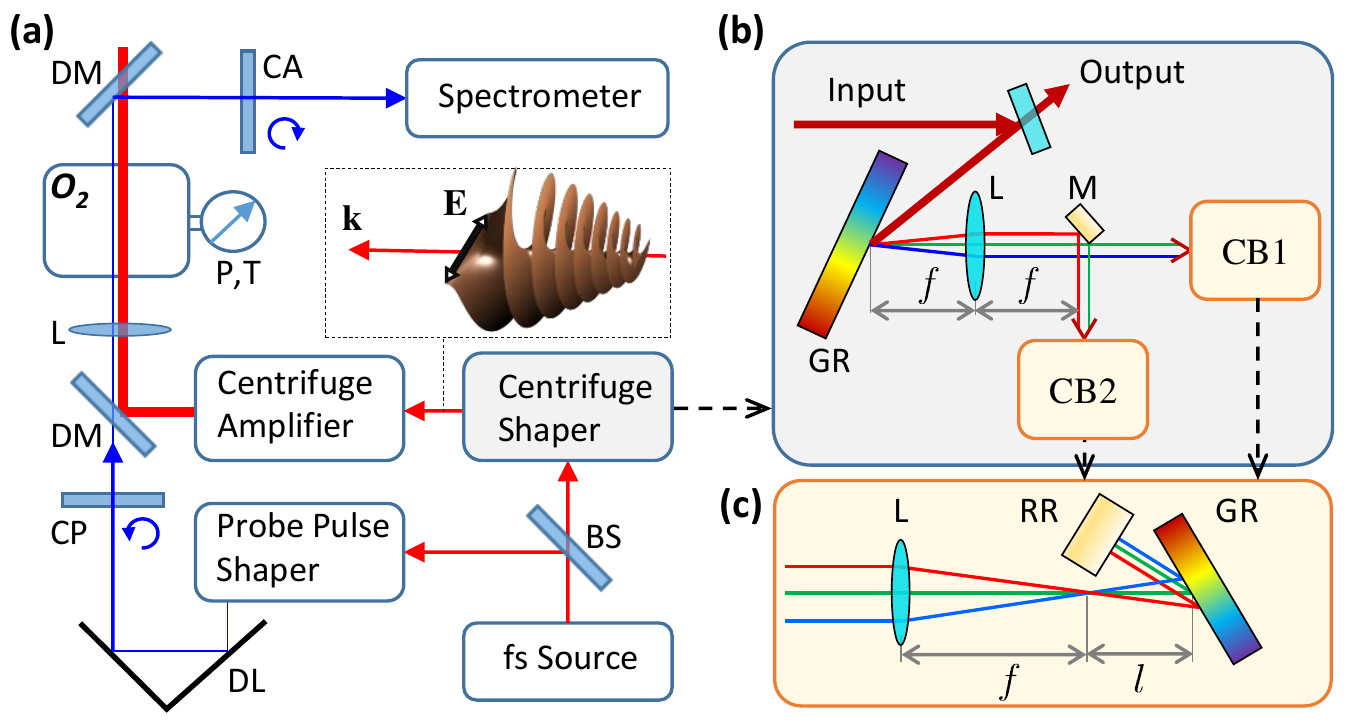}
\caption{(\textbf{a}): Experimental set up. BS: beam splitter, DM: dichroic mirror, CP/CA: circular polarizer/analyzer, DL: delay line, L: lens. `O$_2$' marks the pressure chamber filled with oxygen gas under pressure $P$ and temperature $T$. An optical centrifuge field is illustrated above the centrifuge shaper with \textbf{k} being the propagation direction and \textbf{E} the vector of linear polarization undergoing an accelerated rotation. (\textbf{b}): Centrifuge shaper. GR: grating, M: pick-off mirror in the Fourier plane of lens L of focal length $f$, CB1 and CB2: two ``chirp boxes'', schematically shown in panel (\textbf{c}), where RR denotes a retro-reflector and length $l$ controls the applied frequency chirp.}
\label{Fig-Setup}
\end{figure}

The experimental setup is similar to that used in our original demonstration of molecular super rotors\cite{Korobenko14a}. As shown in Fig.\ref{Fig-Setup}(\textbf{a}), a beam of femtosecond pulses from a regenerative chirped pulse amplifier (spectral full width at half maximum (FWHM) of 30 nm, chirped pulse length 140 ps, FWHM) is split in two parts. One part is sent to the ``centrifuge shaper'' which converts the input laser field into the field of an optical centrifuge. Shown in Fig.\ref{Fig-Setup}(\textbf{b}), the centrifuge shaper is implemented according to the original recipe of Karczmarek \textit{et al.} \cite{Karczmarek99, Villeneuve00}. It consists of a grating-lens pair which disperses the spectral components of the pulse in space. A pick-off mirror, placed in the Fourier plane of the lens, splits the beam in two parts which are sent to two ``chirp boxes'', CB1 and CB2. CB1 preserves the frequency chirp of the input pulse, whereas CB2 changes it to the chirp of an opposite sign and same magnitude. The centrifuge shaper is followed by a home built Ti:Sapphire multi-pass amplifier boosting the energy of each chirped pulse up to 30 mJ. The two amplified pulses are then circularly polarized in an opposite direction with respect to one another and spatially re-combined into a single beam. Optical interference of the two oppositely chirped and circularly polarized components produces the field of an optical centrifuge, schematically illustrated in the inset to Fig.\ref{Fig-Setup}(\textbf{a}). Centrifuge pulses are about 100 ps long, and their linear polarization undergoes an accelerated rotation, reaching the angular frequency of 10 THz by the end of the pulse.  The second (probe) beam passes through the standard $4f$ Fourier pulse shaper employed for narrowing the spectral width of probe pulses down to 3.75 cm$^{-1}$ (FWHM). The central wavelength of probe pulses is shifted to 398 nm by means of the frequency doubling in a nonlinear BaB$_2$O$_4$ crystal.
\begin{figure}[tb]
\centering
\includegraphics[width=.9\columnwidth]{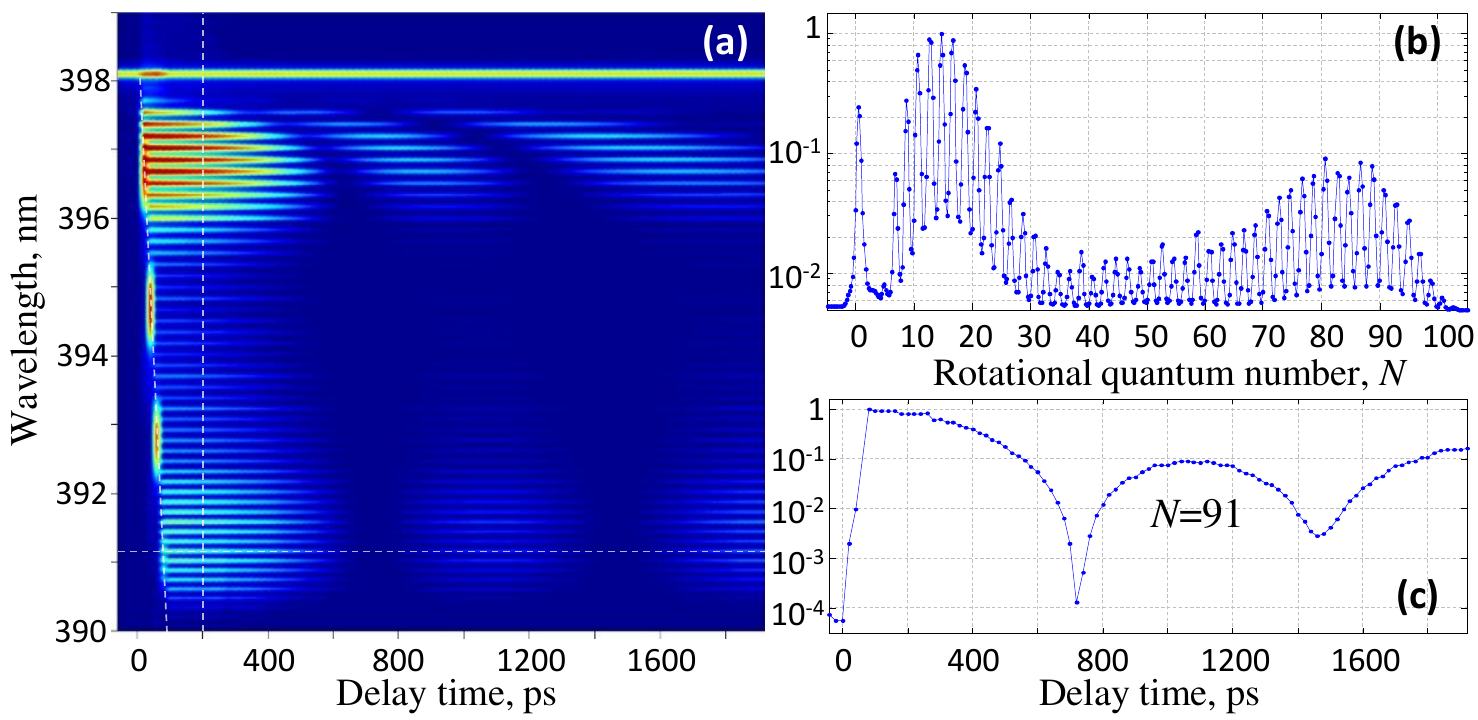}
\caption{(\textbf{a}): Experimentally detected Raman spectrogram of centrifuged oxygen showing the rotational Raman spectrum as a function of the time delay between the beginning of the centrifuge pulse and the arrival of the probe pulse. Color coding is used to reflect the signal strength in logarithmic scale. Tilted white dashed line marks the linearly increasing Raman shift due to the accelerated rotation of molecules inside the 100 ps long centrifuge pulse. (\textbf{b}): Cross-section of the two-dimensional spectrogram at the delay of 200 ps (vertical dashed line in \textbf{a}), showing an ultra-broad rotational wave packet created by the optical centrifuge. (\textbf{c}): Spin-rotation oscillations of the $N=91$ Raman line (horizontal dashed line in \textbf{a}). Note logarithmic scale in all panels.}
\label{Fig-Spectrogram}
\end{figure}

As demonstrated in our previous work\cite{Korobenko14a}, the centrifuge-induced coherence between the states $|J,m=J\rangle$ and $|J+2,m=J+2\rangle$ (where $m$ is the projection of $\vec{J}$ on the propagation direction of the centrifuge field) results in the Raman frequency shift of the probe field. From the selection rule $\Delta m=2$ and the conservation of angular momentum, it follows that the Raman sideband of a circularly polarized probe is also circularly polarized, but with an opposite handedness. Due to this change of polarization, the strong background of the input probe light can be efficiently suppressed by means of a circular analyzer, orthogonal to the input circular polarizer (CA and CP, respectively, in Fig.\ref{Fig-Setup}(\textbf{a})).

The Raman spectrum of the probe pulses scattered off the centrifuged molecules is measured with an f/4.8 spectrometer equipped with a 2400 lines/mm grating as a function of the probe delay relative to the centrifuge. An example of the experimentally detected Raman spectrogram is shown in Fig.\ref{Fig-Spectrogram}(\textbf{a}). It reflects the accelerated spinning of molecules inside the centrifuge during the first 100 ps (marked by a tilted dashed white line). While spinning up, the molecules are ``leaking'' from the centrifuge, producing a whole series of Raman sidebands - a set of horizontal lines shifted from the probe central wavelength of 398 nm. Narrow probe bandwidth enables us to resolve individual rotational states and make an easy assignment of the rotational quantum numbers to the observed spectral lines. This is demonstrated by the Raman spectrum taken at $t=200$ ps and shown in Fig.\ref{Fig-Spectrogram}(\textbf{b}). The created wave packet consists of a large number of odd $N$-states, with even $N$'s missing due to the oxygen nuclear spin statistics. Each Raman line undergoes quasi-periodic oscillations due to the interference between the three frequency-unresolved components $S_{1,2,3}(N)$ of the $S(N)$ branch split by the spin-rotation interaction. An example of these spin-rotation oscillations for the $N=91$ Raman line is shown in Fig.\ref{Fig-Spectrogram}(\textbf{c}). The oscillations start at around 100 ps, after the super rotors with the rotational angular momentum of $91\hbar$ have escaped from the centrifuge.

\begin{figure}[tb]
\centering
\includegraphics[width=1\columnwidth]{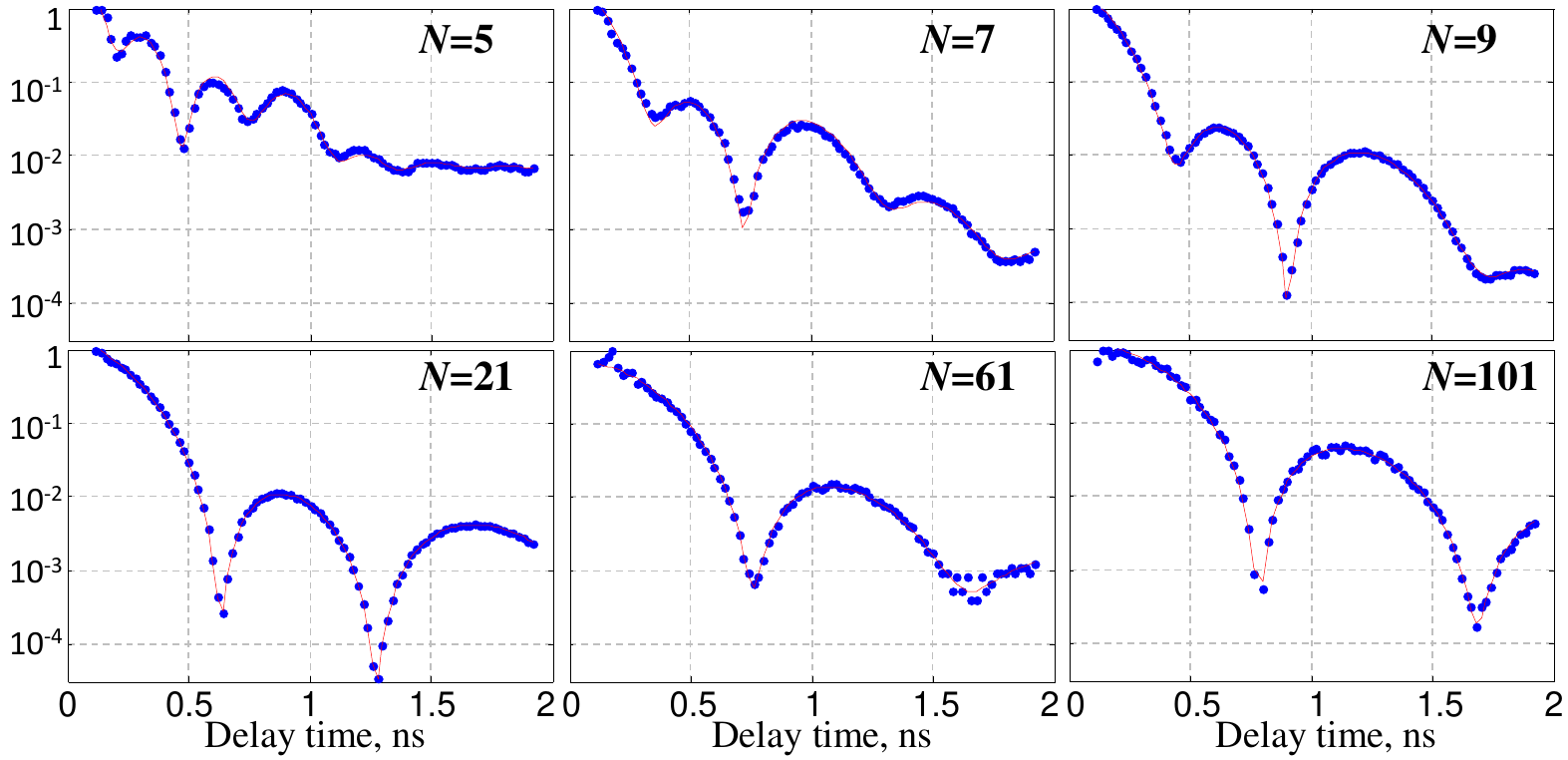}
\caption{The observed data (blue circles, normalized to 1 at $t=100$ ps) and the fit to spin-rotation oscillations (red curves, Eq.\ref{Eq_RamanIntensity}) for six different Raman lines corresponding to the rotational quantum numbers $N=5,7,9,21,61$ and 101. Note logarithmic scale in all panels.}
\label{Fig-Fits}
\end{figure}

The intensity of a Raman line corresponding to the transition between the sates $N$ and $N-2$ can be described as
\begin{equation}\label{Eq_RamanIntensity}
    I_{N}(t) = I_{0} \left| \rho_{N,N-2}(t)\right|^2 e^{-t/\tau_{_N} },
\end{equation}
where $I_{0}$ is determined by a number of time-independent parameters, such as molecular concentration and probe intensity, $\tau_{_N} $ is the collisional decay time constant and $\rho _{N,N-2}(t)$ is the centrifuge induced coherence between the corresponding rotational states. As discussed above, at $N\geqslant5$, the latter consists of three main frequency components corresponding to the three $S$ branch transitions (see Fig.\ref{Fig-Splitting}),
\begin{equation}\label{Eq_Coherence}
    \rho _{N,N-2}(t)=\sum_{k=1,2,3} a_{k} e^{i\Omega _{k}(N)(t-t_{0})},
\end{equation}
with amplitudes $a_{k}$ and frequencies $\Omega _{k}(N)$. Time $t_{0}$ ($0<t_{0}<100$ ps) represents the release time of the corresponding rotational state from the centrifuge. For any $N$, the three frequencies are simply $\Omega _{k}(N)=\left[F_{k}(N)-F_{k}(N-2)\right]/h$, where $F_{k}(N)$ are the well known spin-rotational energies of oxygen\cite{HerzbergBook} and $h$ is the Plank's constant. After normalizing each measured Raman line to 1 at $t=100$ ps (i.e. shortly after the end of the centrifuge pulse), we fit the theoretical expression to the observed signals using the following five fitting parameters $\left\{ a_{1},a_{2},a_{3}, t_{0}, \tau_{_N} \right\}$. As demonstrated by a few examples in Fig.\ref{Fig-Fits}, the oscillatory behavior of our experimental data is well described by Eq.\ref{Eq_RamanIntensity} over the whole range of angular momentum accessed by the centrifuge, from $N=5$ to $N=109$. Note that the weaker the line (e.g. $N=5$) the smaller the dynamic range, ultimately determined by the sensitivity of our detector.

\begin{figure}[tb]
\centering
\includegraphics[width=.85\columnwidth]{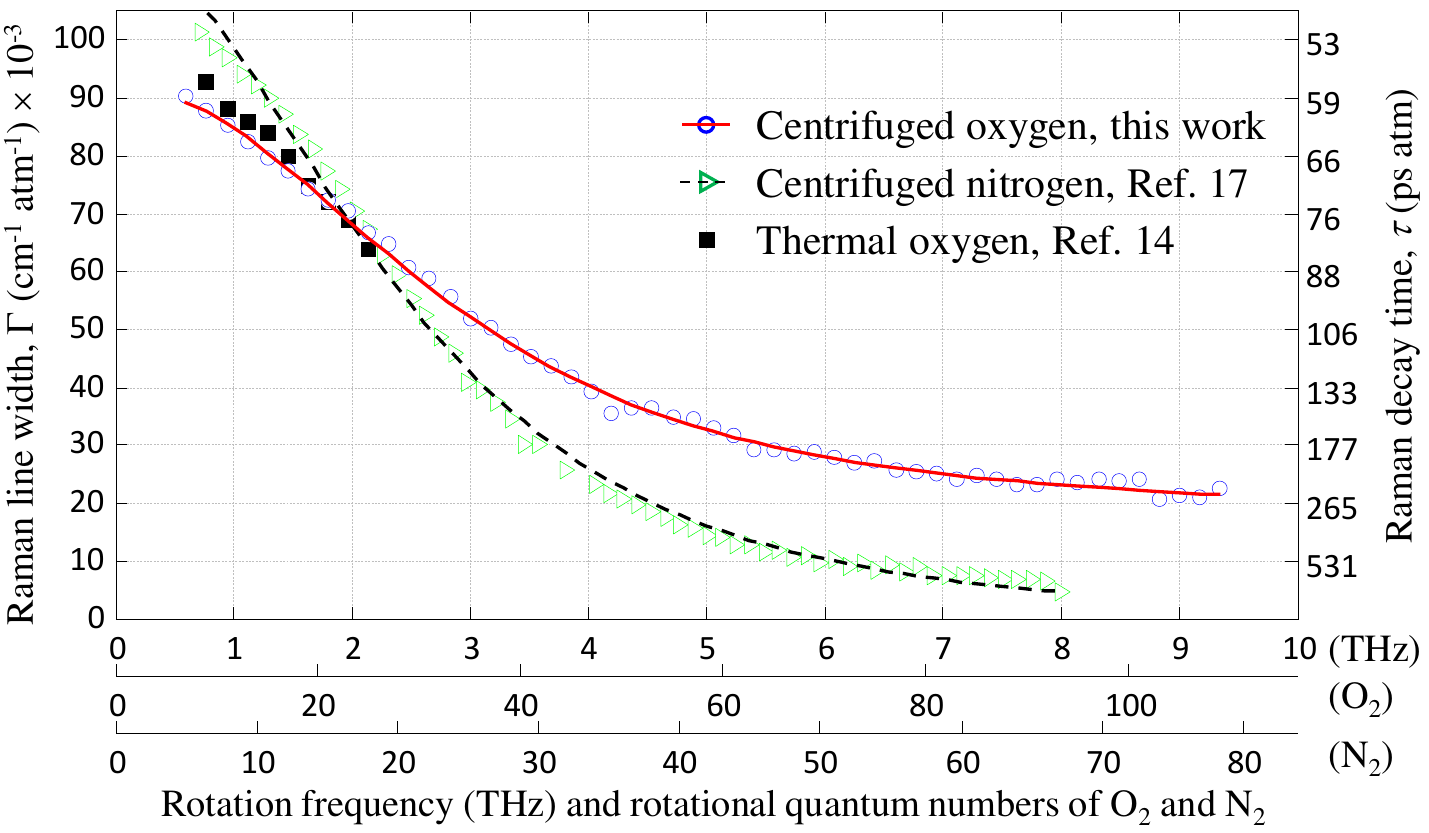}
\caption{The decay rate of rotational coherence in oxygen (blue circles, this work) and nitrogen (green triangles \cite{Milner14a}) as a function of the frequency of molecular rotation. For convenience, rotational quantum numbers of \Otwo and \Ntwo are shown below the frequency axis, and the decay times $\tau_{_N} $ are shown on the right vertical axis. Black squares depict the data from \cite{Miller11}, where the rotational decay has been studied in thermal oxygen. Solid red (dashed black) curve shows the result of the energy corrected sudden (ECS) model of rotational relaxation applied to oxygen (nitrogen) and discussed in text.}
\label{Fig-Comparison}
\end{figure}

From the fitting procedure described above, we retrieve the time constant $\tau_{_N} $ of the collision-induced exponential decay of rotational coherence. For the slower rotating molecules, the coherence life time is shorter than for the faster rotors. This is shown with blue circles in Fig.\ref{Fig-Comparison}, where the decay rate (expressed in the units of Raman line width, $\Gamma_N \equiv \left[ 2\pi c \tau_{_N} \right]^{-1} $, with $c$ being the speed of light in vacuum) is plotted as a function of the angular frequency of molecular rotation. Black squares depict the data from \cite{Miller11} obtained in a thermal ensemble of oxygen molecules at room temperature (hence, $N\leqslant25$) and showing satisfactory agreement with our results at low $N$'s.

It has been suggested that inelastic collisions accompanied by the rotational energy transfer are the main contributors to the rotational decoherence studied in this work\cite{Strekalov00}. Hence, the increasing coherence life time with increasing angular momentum is well expected from the scaling laws of the collisional energy transfer\cite{Polanyi72}. In the popular energy corrected sudden scaling law\cite{Depristo79}, the rate constant  $\gamma^{\textsc{esc}}_{N,N'}$ for the transition from $N$ to $N'$ is described in terms of the basis rate constants $\gamma_{L,0}$ in the following way:
\begin{eqnarray}\label{Eq_ECS}
    \gamma^{\textsc{esc}}_{N,N'} = (2N'+1) \; \exp \left( \frac{E_{N}-E_{N_>}}{k_{B}T} \right) \times \nonumber \\ \sum_{L} \left( \begin{array}{ccc} N & N' & L \\ 0 & 0 & 0 \end{array} \right)^{2} (2L+1) \frac{\Omega_{l_c,v_c} (N)}{\Omega_{l_c,v_c} (L)} \gamma_{L,0},
\end{eqnarray}
where $E_{N}$ is the rotational energy (with $N_>$ denoting the greater of $N$ and $N'$), (:::) is the Wigner $3J$ symbol, and $\Omega _{l_c,v_c}(N)$ is an adiabaticity correction factor. The latter is expressed through an adiabaticity parameter $a_{_N}$ corresponding to the angle, by which a molecule rotates during the collision process,
\begin{equation}\label{Eq_AdiabatictyFactor}
    \Omega_{l_c,v_c}(N) \equiv \left( \frac{1}{1+ a_{_N}^2/6} \right) ^{-2},
\end{equation}
where $a_{_N} \equiv \omega _{_N} \tau_c = \omega _{_N} l_c / v_c$, $\omega _{_N}$ is the frequency of molecular rotation, $\tau_c$ is the collision time, $l_c$ is a characteristic interaction length and $v_c$ is the mean relative velocity between the collision partners. When the period of molecular rotation becomes comparable with, and even shorter than, the time of a single collision ($a_{_N} \geqslant \pi$), molecular interactions become more adiabatic and the rotational coherence more robust with respect to collisions. The basis rates $\gamma_{L,0}$ reflect the probability of changing the angular momentum from $L$ to 0 in a single collision event. They are typically assumed to decrease with $L$ according to the power law,
\begin{equation}\label{Eq_BasisRates}
    \gamma_{L,0} = \frac{A}{\left[ L(L+1) \right]^\alpha}.
\end{equation}
The experimentally observed decay rate of any $N \rightarrow (N-2)$ Raman line can be calculated as a sum over all allowed decay channels for both participating levels, i.e.
\begin{equation}\label{Eq_DecayRate}
    \Gamma _{N}=\sum_{N'} \left( \gamma^{\textsc{esc}}_{N,N'} + \gamma^{\textsc{esc}}_{N-2,N'}\right) \approx 2 \sum_{N'} \gamma^{\textsc{esc}}_{N,N'}.
\end{equation}
Red solid curve in Fig.\ref{Fig-Comparison} shows the result of fitting Eq.\ref{Eq_ECS} to our data using the three fitting parameters of the ECS model: $A, \alpha $ and $l_c$. Good fit is achieved with the values $A=44.6 \times 10^{-3}$ cm$^{-1}$ atm$^{-1}$, $\alpha=1.12$ and  $l_c = 0.57$ \AA\, which are in reasonable agreement with the previously reported values\cite{Thumann97}.

It is instructive to compare these results with our recent study of rotational decoherence in centrifuged nitrogen\cite{Milner14a}. Shown with green triangles in Fig.\ref{Fig-Comparison}, the decay rates in nitrogen are noticeably lower at high values of angular momentum. The best fit is provided by $A=219 \times 10^{-3}$ cm$^{-1}$ atm$^{-1}$, $\alpha=1.62$ and  $l_c = 0.62$ \AA (dashed black curve). The main difference with respect to oxygen is in the higher exponent $\alpha$ which, according to Eq.\ref{Eq_BasisRates}, corresponds to the lower contribution of rotational transitions with large change of molecular angular momentum (this is also the reason for the increased value of $A$).

Different scaling of the basis rates $\gamma_{L,0}$ with $L$ may point at a different mechanism of rotational decoherence in magnetic (O$_{2}$) and non-magnetic (N$_{2}$) molecules. Alternatively, faster decoherence of oxygen rotation may stem from the higher concentration of O$_{2}$ super rotors with respect to the concentration of the centrifuged N$_{2}$. Though not yet explained, this empirically found difference may result in the increasing local gas temperature and correspondingly higher rates of rotational energy transfer. Quantitative studies of this phenomenon are underway.

In summary, we have observed the spin-rotation dynamics in the gas of optically centrifuged oxygen molecules. Because of the interaction between the rotational magnetic moment and the electronic spin, molecules with the different spin orientation with respect to their total angular momentum rotate with a slightly different frequency. Frequency beating of the three spin components results in the spin-rotation oscillations detected in this work. Time-resolved characterization of the SR oscillations may prove useful for creating ensembles of simultaneously spatially-aligned and spin-polarized molecules. The decay of the spin-rotation coherence due to collisions has been quantified and explained in the framework of the energy corrected scaling law of rotational relaxation.

\ack{This work has been supported by the CFI, BCKDF and NSERC.}

\section*{References}
\bibliographystyle{unsrt}

\end{document}